\documentclass[manuscript]{acmart}

\AtBeginDocument{%
  }

\copyrightyear{2024}
\acmYear{2024}
\setcopyright{rightsretained}
\acmConference[CHI '24]{Proceedings of the CHI Conference on Human Factors in Computing Systems}{May 11--16, 2024}{Honolulu, HI, USA}
\acmBooktitle{Proceedings of the CHI Conference on Human Factors in Computing Systems (CHI '24), May 11--16, 2024, Honolulu, HI, USA}
\acmDOI{10.1145/3613904.3642116}
\acmISBN{979-8-4007-0330-0/24/05}

\usepackage{graphics}
\usepackage[flushleft]{threeparttable}
\usepackage{multirow}
\usepackage{graphicx}
\usepackage{array}
\usepackage{longtable}
\usepackage{setspace}
\usepackage{booktabs}
\usepackage{xcolor,colortbl}
\usepackage{url}

\newcommand{\totalAIAAIC}{1,049}
\newcommand{\totalcase}{321}

\begin{document}


\title[A Taxonomy of AI Privacy Risks]{Deepfakes, Phrenology, Surveillance, and More! A Taxonomy of AI Privacy Risks}


\author{Hao-Ping (Hank) Lee}
\email{haopingl@cs.cmu.edu}
\affiliation{%
  \institution{Carnegie Mellon University}
  \city{Pittsburgh}
  \state{PA}
  \country{United States}
}

\author{Yu-Ju Yang}
\email{yujuy@andrew.cmu.edu}
\affiliation{%
  \institution{Carnegie Mellon University}
  \city{Pittsburgh}
  \state{PA}
  \country{United States}
}

\author{Thomas Serban von Davier}
\email{thomas.von.davier@cs.ox.ac.uk}
\affiliation{%
  \institution{University of Oxford}
  \city{Oxford}
  \country{United Kingdom}
}

\author{Jodi Forlizzi}
\email{forlizzi@cs.cmu.edu}
\affiliation{%
  \institution{Carnegie Mellon University}
  \city{Pittsburgh}
  \state{PA}
  \country{United States}
}

\author{Sauvik Das}
\email{sauvik@cmu.edu}
\affiliation{%
  \institution{Carnegie Mellon University}
  \city{Pittsburgh}
  \state{PA}
  \country{United States}
}

\renewcommand{\shortauthors}{Lee et al.}


\begin{abstract}
Privacy is a key principle for developing ethical AI technologies, but how does including AI technologies in products and services change privacy risks?
We constructed a taxonomy of AI privacy risks by analyzing \totalcase{} documented AI privacy incidents.
We codified how the unique capabilities and requirements of AI technologies described in those incidents generated new privacy risks, exacerbated known ones, or otherwise did not meaningfully alter the risk.
We present 12 high-level privacy risks that AI technologies either newly created (e.g., exposure risks from deepfake pornography) or exacerbated (e.g., surveillance risks from collecting training data).
One upshot of our work is that incorporating AI technologies into a product can alter the privacy risks it entails.
Yet, current approaches to privacy-preserving AI/ML (e.g., federated learning, differential privacy, checklists) only address a subset of the privacy risks arising from the capabilities and data requirements of AI.

\end{abstract}



\begin{teaserfigure}
  \includegraphics[width=\textwidth]{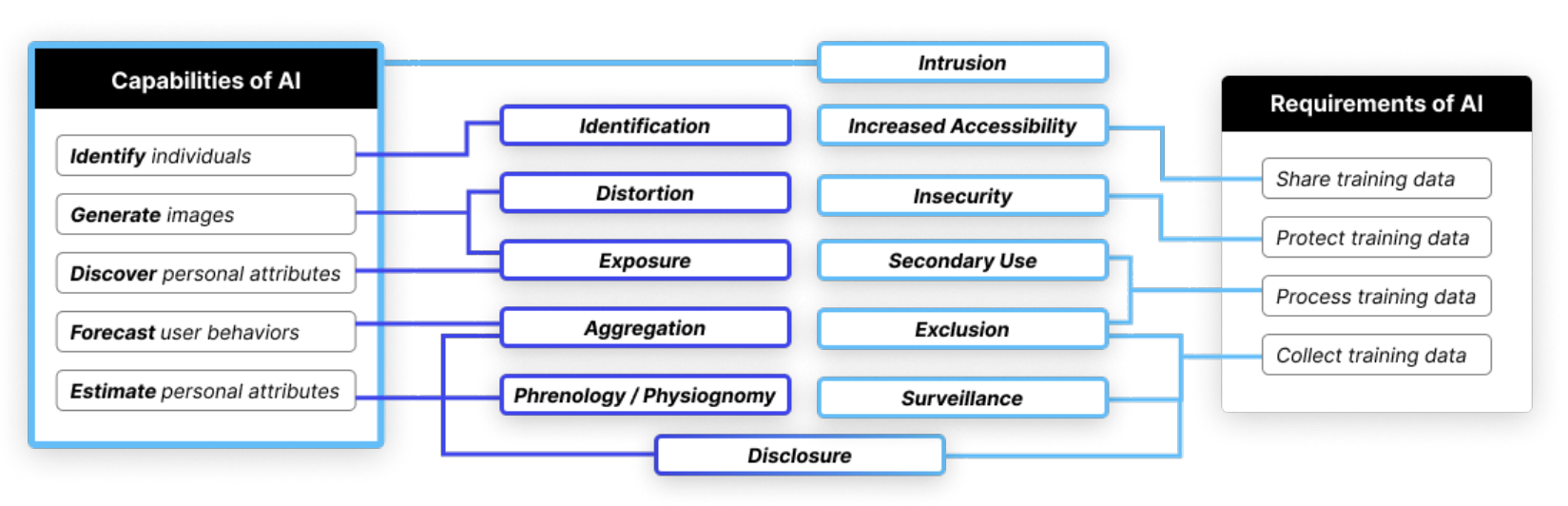}
  \caption{
  \textcolor{black}{We identify 12 privacy risks that the unique capabilities and/or requirements of AI can entail. For example, the capabilities of AI create new risks (purple) of identification, distortion, physiognomy, and unwanted disclosure; the data requirements of AI can exacerbate risks (light blue) of surveillance, exclusion, secondary use, and data breaches owing to insecurity.}}
  \label{fig:overview}
\end{teaserfigure}

\maketitle

\section{Introduction}
In January 2020, privacy journalist Kashmir Hill published an article in the New York Times describing Clearview.AI --- a company that purports to help U.S. law enforcement match photos of unknown people to their online presence through a facial recognition model trained by scraping millions of publicly available face images online \cite{hill_secretive_2020}. In 2021, police departments in many different U.S. cities were reported to have used Clearview.AI to identify individuals, including Black Lives Matter protesters \cite{rihl_emails_2021}. In 2022, a California-based artist found that photos she thought to be in her private medical record were included, without her knowledge or consent, in the LAION training dataset that has been used to train Stable Diffusion and Google Imagen \cite{edwards_artist_2022}. The artist has a rare medical condition that she preferred to keep private, and expressed concern about the abusive potential of generative AI technologies having access to her photos. In January 2023, Twitch streamer QTCinderella made an emphatic plea to her followers on Twitter to stop spreading links to an illicit website hosting AI-generated deepfake pornography of her and other women influencers. ``Being seen `naked' against your will should NOT BE A PART OF THIS JOB'' \cite{qtcinderella_qtcinderella_i_2023}.

These examples illuminate the unique privacy risks posed by AI technologies, prompting the foundational research question we ask in this work: \textbf{How do modern advances in AI and ML change the privacy risks of a product or service?} To answer this question, we introduce a taxonomy of AI privacy risks, grounded in an analysis of \totalcase{} privacy-relevant incidents that resulted from AI products and services, sourced from an AI incidents database \cite{pownall_ai_2023}, much like the ones described above.
This work is important for at least two reasons. First, people are concerned about how AI can affect their privacy: 
a 2021 survey with around 10,000 participants from ten countries found that roughly half of the respondents believed that AI would result in ``less privacy'' in the future, citing concerns around large-scale collection of personal data, consent, and surveillance \cite{kelley_there_2023}.
Second, while privacy is one of the five most commonly cited principles for the development of ethical AI technologies \cite{jobin_global_2019}, we do not yet have a systematic understanding of if and how modern advances in AI change the privacy risks entailed by products and services.

While AI and ML technologies have vastly expanded in capability \cite{yildirim_how_2022}, there is simultaneously a great deal of hype about what these technologies can and cannot do, making it difficult to separate real risks from speculative ones \cite{kapoor_ai_2023}. Thus, it can be difficult for today's practitioners who develop AI-inclusive products and services to understand how their use of AI technologies might entail or exacerbate practical privacy risks \cite{yurrita_towards_2022}. Prior work shows this difficulty to be true: in an interview with 35 AI practitioners, Lee et al. found that participants had relatively low awareness of privacy risks unique to or exacerbated by AI, and had little incentive to and support in addressing these risks \cite{lee2024}.

AI and privacy both existed long before modern dialogues around the role of privacy in ethical AI development. To understand what modern advances in AI \textit{change} about privacy, we needed a suitable baseline for privacy risk as it was understood before these advances.
To that end, we used Solove's highly-cited and well-known taxonomy of privacy from 2006 as a baseline \cite{solove_taxonomy_2006}. Solove's taxonomy was proposed well before modern advances in AI became mainstream in product design, and remains relevant and influential to this day. Yet, Solove's taxonomy is intentionally broad and technology-agnostic --- a useful attribute in the legal and regulatory contexts for which it was developed, but less helpful in prescribing specific mitigations for product designers and developers.

To ground our analysis on real and practical risks, we sourced case studies from a database indexing real AI incidents documented by journalists --- the AI, Algorithmic, and Automation Incident and Controversy (AIAAIC) repository \cite{pownall_ai_2023}. 
We sourced \totalcase{} case studies from the AIAAIC repository in which real AI products resulted in lived privacy risks. We next systematically analyzed whether and how the capabilities and/or requirements of the AI technology described in the incident either (i) \textit{created} a new instantiation of a privacy risk described in Solove's original taxonomy or an entirely new category of risk, (ii) \textit{exacerbated} a privacy risk that was already captured by Solove's taxonomy, or (iii) \textit{did not change} the privacy risk described in the incident relative to at least one of the risks described in Solove's taxonomy. 


The result is our taxonomy of AI privacy risks (see Figure~\ref{fig:overview}). Our taxonomy illustrates how the unique capabilities of AI --- e.g., the ability to \textit{recommend} courses of action, \textit{infer} users' interests and attributes, and \textit{detect} rare or anomalous events \cite{google2019people} --- resulted in both new instantiations of existing categories of risk in Solove's taxonomy as well as one entirely new category of privacy risk. For example, we found that the ability of AI technologies to generate human-like media resulted in new types of exposure risks (e.g., the generation of deepfake pornography \cite{noauthor_deepfake_2021}), while the ability for AI to learn arbitrary classification functions led to a new category of privacy risk: phrenology/physiognomy (e.g., the belief that AI can be used to automatically detect things like sexual orientation from physical attributes \cite{levin_lgbt_2017}. 
Our taxonomy also captures how the data and infrastructural requirements of AI exacerbated privacy risks already captured in Solove's taxonomy.
For example, since facial recognition classifiers require tremendous amounts of face data, they can exacerbate surveillance risks by encouraging uncritical data collection practices such as collecting face scans in airports \cite{fassler_south_2021}.

We discuss how existing approaches to privacy-preserving AI and machine learning, such as differential privacy and federated machine learning, only account for a subset of these risks, highlighting the need for new tools, artifacts, and resources that aid practitioners in negotiating the utility-intrusiveness trade-off of AI-powered products and services. Finally, we outline how this taxonomy can be used to create tools that help educate practitioners, and as a repository of shared knowledge regarding AI privacy risks and design processes to mitigate against those risks.

\section{Background and Related Work}

\subsection{Human-centered AI}

AI technologies are here to stay. New AI technologies are constantly created and evolving, and so are their harms to individuals and society \cite{pownall_ai_2023}. Advertising in which users' interests are inferred from their behaviors online to target them with relevant advertisements fuels a multi-trillion dollar industry that has been referred to as ``surveillance capitalism.'' \cite{Zuboff2019} Users find these ads both ``smart'' and ``scary'' \cite{ur_smart_2012}. Beyond attitudes, recent work has further shown that these advertisements result in many real, lived harms --- ranging from psychological distress to traumatization \cite{Wu2023}.
As AI technologies improve, we see new uses of these technologies to make spurious predictions about individuals and their behavior, portending a new age of AI-facilitated phrenology and physiognomy: e.g., through the use of profile images to predict things like sexual orientation \cite{wang2018deep} and ``criminality'' \cite{wu2016automated}.

In response to the potentially detrimental effects of unchecked AI on society, 
there is growing discussion on how AI technologies' benefits can be ensured and their potential harms mitigated \cite{desouza_designing_2020}. Human-centered AI (HAI) is a term commonly used to center human needs and to describe the ethical decision-making that informs AI design \cite{tigard_responsible_2021,sambasivan_toward_2018, barredo_arrieta_explainable_2020}. In recent years, Human-computer Interaction (HCI) researchers and AI practitioners have created a body of work to provide guidelines for HAI (see Hagendorff \cite{hagendorff_ethics_2020} for an extensive evaluation).

Case studies on implementing HAI guidelines reveal stakeholders' struggle with concepts of privacy and fairness \cite{vyhmeister_lessons_2022, jobin_global_2019, fjeld_principled_2020}. This paper focuses on privacy, as significant research has previously attempted to define and measure fairness in AI \cite{chester_balancing_2020,bellamy_ai_2018}. To understand the potential privacy risks of AI technologies, the negative impacts of past implementations must be considered \cite{mikalef_thinking_2022}. This method of looking at the ``dark side'' of technologies can reveal the potential risks of future technology concepts by reflecting on past harms and has been successfully used to analyze dark patterns in GUIs and consider how software agents may impact user autonomy \cite{mathur_dark_2019, friedman_software_1997}.

We present a novel taxonomy of AI privacy risks to further develop what it means to design for privacy in human-centered AI. This taxonomy aims to provide AI practitioners with tools and a shared language to foreground end-user privacy discussions in the design and development process. 

\subsection{Prior privacy taxonomies and concepts}
\label{sec:prior privacy taxonomies}






AI is unique in its capacity for high-powered decision-making. Unlike traditional tools, AI systems demand copious amounts of data to refine and enhance outputs \cite{Xu2021ArtificialIntelligence}. However, the data often originates from individuals, giving rise to pressing concerns about privacy and safety \cite{Szegedy2013}. Therefore, input from various sectors is needed, along with comprehensive strategies for responsible development and deployment.

Ensuring privacy and AI safety has been addressed from various angles. Many approaches build upon the seminal work in privacy preservation pioneered by Shokri and Shmatikov \cite{Shokri2015}. Other research documents challenges \cite{Liu2020}, especially within the realm of deep learning techniques \cite{Boulemtafes2020, Liu2021}. Furthermore, a spectrum of cybersecurity threats looms over any AI system striving to safeguard the privacy of its users and data providers \cite{Oseni2020}.

While these works address the task of documenting potential privacy challenges and vulnerabilities within AI systems, they often focus on specific aspects rather than taking a holistic view \cite{Shahriar2023}. When researchers examine the full AI ``life-cycle,'' it is typically aimed at promoting and ensuring trust and assurance within AI, rather than concentrating on the initial privacy concerns that precipitated distrust \cite{Ashmore2021, wickramasinghe_trustworthy_2020}.

Shahriar et al. offer four categorizations of privacy risks along with a relevant list of strategies applicable throughout the design, development, and deployment phases of an AI system \cite{Shahriar2023}: (1) the risk of identification, (2) the risk of inaccurate decisions, (3) the risk of non-transparent AI, and (4) the risk of non-compliance with regulations. Their categorizations provide effective catch-alls for various potential risks. However, like other recent frameworks (see \cite{Warford2022}), the approach of categorizing strategies and techniques by privacy risks involves a degree of theoretical dangers outlined in research case studies and previous surveys rather than proven, reported, and documented privacy risks. Moreover, these taxonomies do not consider what AI technologies \textit{change} about privacy risks relative to notions of privacy prior to modern advances (e.g., the creation and use of deepfake techniques).

It is crucial to turn to the literature on privacy law to address this gap and provide a more holistic and inclusive understanding of AI privacy risks as they manifest worldwide. Solove's work represents the progress within legal discussions and the judicial system to address taxonomies of different types of privacy risks \cite{solove_taxonomy_2006}. Solove's taxonomy offered a comprehensive classification of different types of privacy intrusions (i.e., intrusions associated with information collection, information
processing, information dissemination, and invasion) as seen in the legal field. It was previously used in security research to explore users' personal attitudes and behaviors regarding privacy issues \cite{KOKOLAKIS2017122, acquisti2015privacy}. However, unlike this paper, the previous research did not look to apply or change the taxonomy to the new and emerging challenges of realized privacy intrusions. 

Solove and colleagues built on their work by defining what qualifies as a ``harm'' and how modern technology challenges these traditional distinctions \cite{Citron2022}. Nevertheless, this taxonomy does not possess the AI-specific focus of Shahriar et al.'s research. Our paper intends to bridge the divide between these two bodies of work. By adopting an AI-centric approach to codify realized privacy risks, our paper introduces a distinctive taxonomy for evaluating and ultimately addressing privacy risks specific to the ``life-cycle'' of AI systems.

\subsection{Creating a privacy taxonomy}
A robust taxonomy can provide AI practitioners with guidance and structure during the design and development process. Taxonomies provide an organizational hierarchy of information, classifying information into distinct categories \cite{chilton_cascade_2013}. However, developing an effective privacy taxonomy is a challenge many researchers have undertaken with limited success \cite{usman_taxonomies_2017}. Two characteristics make the development of a privacy taxonomy challenging.

The first is the inability to agree on any one definition of privacy. Early interpretations consider privacy ``the right to be let alone'' \cite{brandeis_right_1890}. Later, the foundation of modern privacy law was built on an argument calling for individual or group autonomy over the sharing and disseminating of personal information \cite{westin_privacy_1967}. Privacy theory also began to change to consider the dangers of inflexible regulations and the importance of treating privacy as a process rather than a label \cite{altman_environment_1975}. HCI researchers built on this theory to consider applications in practice \cite{palen_unpacking_2003}. More recent work considers privacy in the light of contextual integrity \cite{nissenbaum_privacy_2004}. Other researchers embrace the difficulty of defining privacy as the reason for developing adaptive solutions and classification systems \cite{mulligan_privacy_2016}. In some cases, it has been easier to define privacy within the constraints of a specific field of operation, such as databases \cite{barker_data_2009}. 

The second characteristic that makes creating a universally accepted taxonomy difficult is the need to operationalize the taxonomy in a single domain. For example, there was a push for the taxonomy and approach of privacy by design for ubiquitous computing \cite{goos_privacy_2001}. Similarly, in the field of Robotics, a specific taxonomy was developed to deal with implementing sensor technology \cite{eick_enhancing_2020}. These taxonomies focus on the potential privacy risks the system or technology poses \cite{nunes_taxonomy_2023}. Even taxonomies built on user or societal input rely on perceived risks instead of reported harms resulting from past usage of similar technology \cite{jakobi_taxonomy_2022}. We identified one taxonomy that accounts for previously recorded privacy risks \cite{solove_taxonomy_2006}. However, this work is geared towards lawmakers and legal professionals rather than AI practitioners. It is not built to address specific privacy risks associated with the functionality and design of AI systems.

Similarly, previous HCI research has attempted to provide practitioners with a privacy taxonomy based on end users' experience, raising awareness for physical privacy intrusions \cite{windl2023}. This previous research shows the ability to apply such taxonomies in practical settings. Yet, it does not attempt to handle the more conceptual instances of AI privacy intrusions that may be invisible to end users but are no less impactful.

Therefore, this paper takes the first step to codify patterns of documented privacy risks resulting from AI's capabilities and data requirements. 

\section{Method}
\label{sec:method}
\begin{figure*}[h]
\centering
\includegraphics[width=0.8\columnwidth]{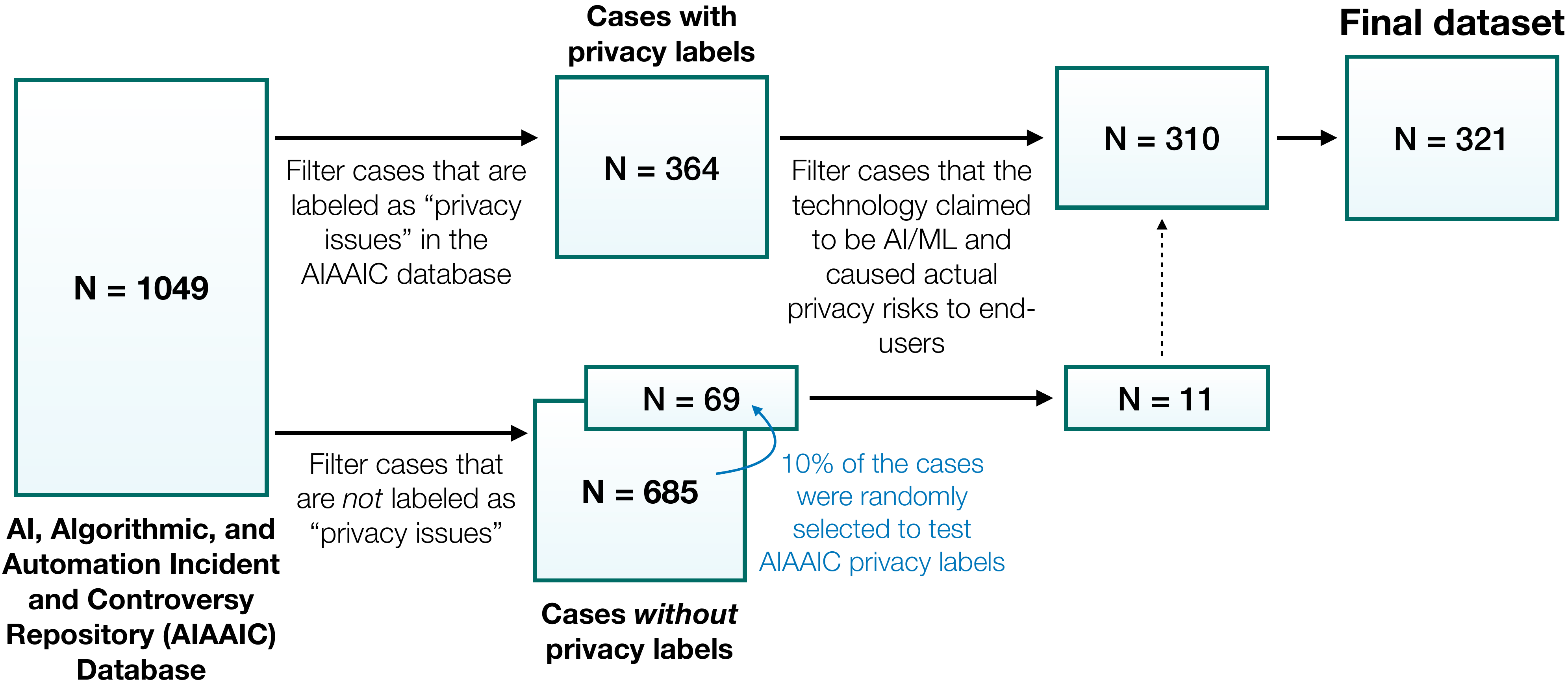}
\caption{We filtered from \totalAIAAIC{} cases from the AIAAIC database and selected cases labeled as ``privacy issues.'' We filtered them down to cases with the technology claimed to be AI/ML that caused actual privacy risks to end-users. We also picked 10\% of the cases without the privacy label from the database and went through the same analysis process. The final dataset comprised a total of \totalcase{} cases.}
\label{fig:method}
\end{figure*}

\subsection{Constructing the Taxonomy of AI Privacy Risks Based on AIAAIC}
We developed a taxonomy of privacy risks exhibited in documented AI privacy incidents by performing a systematic review of case studies. Creating typology and taxonomy by synthesizing real-world incidents has been used more broadly in privacy and security \cite{das_breaking_2018} and in AI ethics \cite{raji_fallacy_2022}. In this paper, we define AI broadly to accommodate the wide range of its capabilities to \textit{``perform tasks or behaviors that a person could reasonably deem to require intelligence if a human were to do it''} \cite{riedl2019human}. 
AI is an umbrella term that encompasses many technologies, and our analysis does as well --- we cover approaches ranging from Machine Learning (e.g., prediction and recommendation algorithms), Natural Language Processing (e.g., large language models), Computer Vision (e.g., facial recognition), and Robotics (e.g., home robots, drones). 
Note that we focus on documented end-user privacy risks of actual AI/ML products rather than speculative risks of general AI/ML concepts. We partly relied on the AI, Algorithmic, and Automation Incident and Controversy Repository (AIAAIC), the largest, up-to-date crowdsourced AI incident database curated by journalism professionals \cite{pownall_ai_2023}. We also surveyed the AI Incident Database (AIID)\footnote{\url{https://incidentdatabase.ai/}}, another public AI incident database, but decided not to use it because the AIAAIC provided good coverage of most of the privacy-related incidents in the AIID\footnote{We randomly selected 33\% (N=50) of the AIID total incidents that contain the keyword ``privacy'' (N=151 as of August 16th, 2023). We manually went through the 50 incidents: 20 were either not AI products (e.g., augmented reality applications, executive orders from the government, policies) or not directly related to privacy (e.g., bias, accuracy), and 17 were already included in our AIAAIC database. The remaining 13 (26\%) were not, but we found similar incidents in the AIAAIC database that were already captured by our taxonomy, e.g., incidents related to surveillance, data breaching, distortion made by deepfake AI, and physical invasion of AI technologies.}.




Out of a database of \totalAIAAIC{} cases\footnote{We took a snapshot of the database on August 16th, 2023}, 364 of them were labeled to involve ``privacy issues''\footnote{The AIAAIC database tags each case with two attributes, \textit{Issue(s)} and \textit{Transparency}, to reflect if a given case raises any privacy concerns from the stakeholders and media.} occurring between 2012 to 2023. 
To ensure that the incidents we analyzed indeed involve AI privacy risks, two coders reviewed the linked resources for \textit{all} 364 cases tagged in the AIAAIC as being privacy-pertinent. Then, the two coders went through an incident-by-incident discussion on whether the reported technology (i) claimed to be inclusive of AI, ML, or otherwise ``algorithmic'' approaches, (ii) was actually deployed to real end-users, and (iii) involved some form of end-user privacy risks and/or compromise, and further filtered down to 310 cases.
We filtered out incidents that did not involve AI technologies (N=21, e.g., virtual-reality applications, data leakage unrelated to the use and development of AI), and incidents that were not associated with end-user privacy risks (N=33, e.g., bias \cite{noauthor_uw-madison_2021}, inaccuracy \cite{francisco_tv_2017}, copyright \cite{gershgorn_github_2021}).


To ensure an adequate sampling strategy, we randomly picked 10\% of the cases \textit{without the privacy label} in the AIAAIC database (69 out of 685). We identified privacy risk(s) in 11 of these cases (15.94\%), and all of the identified risks were found in other cases tagged with the privacy label. Thus, we deemed our analysis had reached saturation. In sum, we analyzed a total of \totalcase{} distinct cases in developing our taxonomy of AI privacy risks (see Figure \ref{fig:method}). 


As our objective was to understand how AI \textit{changes} privacy, and not to re-define what \textit{is} privacy, 
we rooted our analysis on Solove's taxonomy of privacy from 2006 as a baseline \cite{solove_taxonomy_2006} --- a popular conceptualization of privacy risks proposed prior to modern advances in AI/ML.
Our primary analytic goal was to identify if and how AI exacerbates and/or creates privacy risks relative to this taxonomy, because doing so will highlight how modern advances in AI do and do not change notions of privacy risk.
We say that AI \textit{exacerbates} privacy risks when the capabilities and/or requirements of the AI technologies are not the root cause of the privacy risk, but increased its scale, scope, frequency, and/or intensity --- e.g., robust identification even with low-quality images. We say that AI \textit{creates} new privacy risks when the capabilities and/or requirements of the AI technology are fundamental enablers of the privacy risk --- e.g., deepfake pornography. Otherwise, we say that the AI has \textit{not meaningfully changed} the privacy risk described in the incident.


For each incident, we assessed if and how the privacy violations described in the incident related to the unique context, capabilities of, and requirements entailed by the AI technologies described in the incident.
We used an iterative coding process to categorize the privacy risk described in the incident.
First, we created our codebook of different types of privacy risks adapted from the taxonomy proposed by Solove \cite{solove_taxonomy_2006}.
Next, we iteratively updated the definition and scope of Solove's initial set of privacy risks to be more specific to the AI privacy incidents in our dataset.
For example, in our incident database, we observed that \textit{Increased Accessibility} typically manifested as increasing public access to otherwise private or access-controlled data for building AI/ML models (e.g., through the release of public datasets).
We also merged risks when they exclusively co-existed in our analysis.
For example, we found that the Appropriation risk, the use of one's identity to serve the aims and interests of another, always manifested with the Distortion risk, disseminating realistic AI-generated false information about individuals.
While these two categories can theoretically be separable (i.e., one can imagine Distortion without Appropriation or Appropriation without Distortion), to keep our taxonomy grounded on real incidents and not theoretical harms, we merged the two categories into a single \textit{Distortion} category. 
Finally, we found an entirely new type of privacy risk, \textit{Phrenology / Physiognomy}, which is not captured in Solove's initial set of privacy risks. This privacy risk is unique to AI due to its capability to estimate sensitive personal attributes (e.g., sexual orientation, ethnicity) of individuals from their physical attributes (e.g., appearance, voice).

In total, we created a final codebook of 12 operationalizable privacy risk labels for AI technologies, including \textit{Surveillance, Identification, Aggregation, Phrenology / Physiognomy, Secondary Use, Exclusion, Insecurity, Exposure, Distortion, Disclosure, Increased Accessibility}, and \textit{Intrusion} (see Table \ref{tab:taxonomy} and Figure \ref{fig:model}).

\subsection{Qualitative Analysis Procedure}
To summarize our qualitative analysis procedure, the first author iteratively applied the codebook to 132 cases to update and better scope the definition of each privacy risk in active discussion with four other authors and constructed the initial codebook.
Another author joined the coding process when the initial codebook was constructed. This author was trained with the codebook and independently coded the same set of 132 cases.
The codes were then iteratively refined and discussed when disagreements occurred until both authors agreed on all codes in the codebook.
To validate the inter-rater reliability, the two coders then independently coded another 65 cases (20\% of our overall analysis pool; N=321) and reached a high agreement, with Cohen's Kappa larger than 0.8 on every type of risk and averaging 0.94 on all types of risks (see Appendix Table \ref{tab:irr}).
One coder then coded the rest of the 124 cases. The final codebook comprises 12 types of privacy risks that we identified across the corpus of \totalcase{} cases.
In determining whether AI newly created, exacerbated, or not meaningfully changed the privacy risks identified in each incident, the two coders engaged in an incident-by-incident discussion for all 321 incidents concerning the root cause of the privacy intrusions, as well as the role AI played in that root cause. The three themes --- i.e., create, exacerbate, and no meaningful change --- naturally emerged during this process.


\section{Taxonomy of AI Privacy Risks}

\begin{figure}
\centering
\includegraphics[width=0.9\columnwidth]{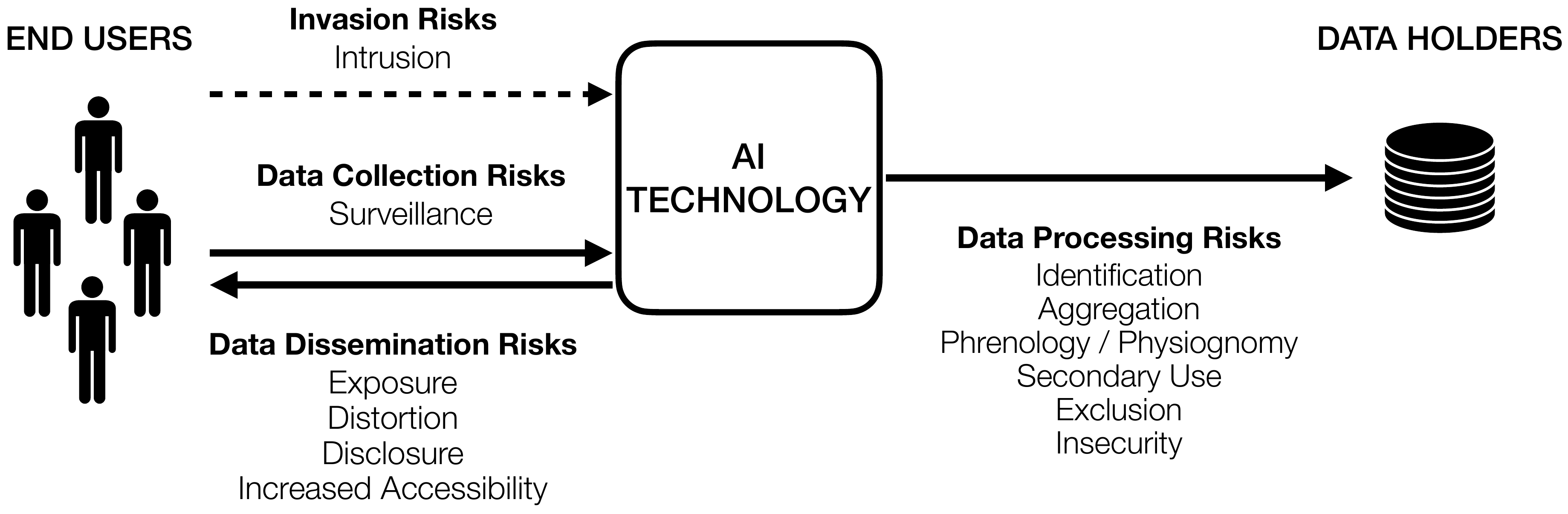}
\caption
      {%
        12 types of privacy risks that AI technologies create and/or exacerbate relate to data collection, data processing, data dissemination, and invasion. The arrows indicate data flow (invasion risks need not involve data, but often do).
      }
\label{fig:model}
\end{figure}

\setlength{\extrarowheight}{0.8ex}
\AtBeginEnvironment{longtable}{\small}
\AtBeginEnvironment{longtable}{\setstretch{0.95}}
\begin{longtable}
{|>{\hspace{0pt}}m{0.205\linewidth}|>{\hspace{0pt}}m{0.285\linewidth}|>{\hspace{0pt}}m{0.43\linewidth}|} 
\caption{Taxonomy of AI Privacy Risks. We found incidents matching 12 distinct, but not mutually exclusive, categories of privacy risk.}
\label{tab:taxonomy}
\\
\hline
\textbf{\textit{Privacy risk}} \cite{solove_taxonomy_2006} & \textbf{\textit{How does AI influence the risk?}} & \textbf{\textit{Examples}} \endfirsthead 
\hline
\multicolumn{3}{|>{\centering\arraybackslash\hspace{0pt}}m{0.92\linewidth}|}{\textbf{\textit{Data Collection Risks}}}
\\
\hline
\textbf{Surveillance}:\par{}\textit{watching, listening to, or recording of an individual’s activities} & AI \textbf{\textit{exacerbates}} surveillance risks by increasing the scale and ubiquity of personal data collected. & A predictive policing platform deployed in Xinjiang, China,``\textit{collects [individual's] information from a variety of sources including CCTV cameras and Wi-Fi sniffers, as well as existing databases of health information, banking records, and family planning history}'' \cite{rajagopalan_china_2018}.\\
\hline
\multicolumn{3}{|>{\centering\arraybackslash\hspace{0pt}}m{0.92\linewidth}|}{\textbf{\textit{Data Processing Risks}}} \\ 
\hline
\textbf{Identification}: \par{}\textit{linking specific data points to an individual’s identity} & AI \textbf{\textit{creates}}\textit{ }create new types of identification risks with respect to scale, latency, robustness, and ubiquity. & Models trained on Simulated Masked Face Recognition Dataset (SMFRD) are capable of identifying persons with a mask on, ``\textit{violating the privacy of those who wish to conceal their face}'' \cite{wiggers_ai_2021}. \\
\hline
\textbf{Aggregation}:\par{}\textit{combining various pieces of data about a person to make inferences beyond what is explicitly captured in those data} & AI \textbf{\textit{creates}} new types of aggregation risks owing to their scale, latency, ubiquity, and their ability to forecast end-user behavior and infer end-user attributes. & ``\textit{The system, called the National Data Analytics Solution (NDAS), uses a combination of AI and statistics to try to assess the risk of someone committing or becoming a victim of gun or knife crime}'' \cite{baraniuk_exclusive_2018}.\\
\hline
\textbf{Phrenology / Physiognomy}:\par{}\textit{inferring personality, social, and emotional attributes about an individual from their physical attributes} & AI \textbf{\textit{creates}}\textit{ }phrenology/physiognomy risks through learning correlations between arbitrary inputs (e.g., images) and outputs (e.g., sexual orientation). & `Gaydar', an AI sexual orientation prediction model, was found to ``distinguish between gay or straight people'' based on their photos \cite{levin_new_2017}. \\ 
\hline
\textbf{Secondary use}:\par{}\textit{the use of personal data collected for one purpose for a different purpose without end-user consent} & AI \textbf{\textit{exacerbates}} secondary use risks by creating new AI capabilities with collected personal data, and (re)creating models from a public dataset. & The Diversity in Faces (DiF) dataset was created to improve the research on fairness and accuracy of artificial intelligence face recognition systems across genders and skin colors, and should not be used for commercial purposes. Nevertheless, Amazon and Microsoft were accused of using the dataset to ``\textit{improve the accuracy of their facial recognition software}'' 
\cite{anne_long_amazon_2021}. \\ 
\hline
\textbf{Exclusion}:\par{}\textit{the failure to provide end-users with notice and control over how their data is being used} & AI \textbf{\textit{exacerbates}} exclusion risks by training on rich personal data without consent. & LAION-5B is a large, openly accessible image-text dataset for training ML models. However, a person found that her private medical photographs were referenced in the public dataset, and suspected that ``\textit{someone stole the image from my deceased doctor's files and it ended up somewhere online, and then it was scraped into this dataset}'' \cite{edwards_artist_2022}. \\ 
\hline
\textbf{Insecurity}:\par{}\textit{carelessness in protecting collected personal data from leaks and improper access due to faulty data storage and data practices} & AI \textbf{\textit{exacerbates}} insecurity risks by introducing new vulnerabilities when incorporating AI and its associated data pipeline in the products. & Lee Luda, a chatbot trained on real-world text conversations, was found to expose the names, nicknames, and home addresses of the users whose data on which it was trained \cite{jang_south_2021}. \\ 
\hline
\multicolumn{3}{|>{\centering\arraybackslash\hspace{0pt}}m{0.92\linewidth}|}{\textbf{\textit{Data Dissemination Risks}}} \\ 
\hline
\textbf{Exposure}:\par{}\textit{revealing sensitive private information that people view as deeply primordial that we have been socialized into concealing} & AI \textbf{\textit{creates}} new types of exposure risks through generative techniques that can reconstruct censored or redacted content; and through exposing inferred sensitive data, preferences, and intentions. & TecoGAN, a deep learning video clarification tool, has been used to clarify censored images of genitalia \cite{montgomery_man_2021}. \\ 
\hline
\textbf{Distortion}:\par{}\textit{disseminating false or misleading information about people} & AI \textbf{\textit{creates}} new types of distortion risks through the generation of realistic fake images and audio that humans have difficulty discerning as fake. & Prime Voice AI, a text-to-voice generator, was misused to create the voices of celebrities to \textit{``make racist remarks about the US House representative''}, and that the AI-generated clips \textit{``run the gamut from harmless, to violent, to transphobic, to homophobic, to racist''} \cite{cox_ai-generated_2023,harrison_startup_2023}.\\ 
\hline
\textbf{Disclosure}:\par{}\textit{revealing and improperly sharing data of individuals} & AI \textbf{\textit{creates}} new types of disclosure risks by inferring additional information beyond what is explicitly captured in the raw data. 
\newline
\vspace{-1mm}
\newline
AI \textbf{\textit{exacerbates}} disclosure risks through sharing personal data to train models. & 
The ``Safe City'' initiative in Myanmar used AI-infused cameras to identify faces and vehicle license plates in public places and alert authorities to individuals with criminal histories \cite{noauthor_myanmar_2021}.
\newline
\vspace{-1mm}
\newline
The UK's National Health Service partnered with Google to share mental health records and HIV diagnoses of 1.6 million patients to develop a model for detecting acute kidney injury \cite{hodson_revealed_2016}. \\ 
\hline
\textbf{Increased Accessibility}:\par{}\textit{making it easier for a wider audience of people to access potentially sensitive information} & AI\textit{ }\textbf{\textit{exacerbates}} the scale of increased accessibility risks via publicizing large-scale datasets that contain personal information, for the use of building and improving AI/ML models. & OkCupid dataset contained personal information such as users' location, demographics, sexual preferences, and drug use, and was uploaded to Open Science Framework to facilitate research on modeling dating behaviors \cite{woollacott_70000_2016}. \\ 
\hline
\multicolumn{3}{|>{\centering\arraybackslash\hspace{0pt}}m{0.92\linewidth}|}{\textbf{\textit{Invasion Risks}}} \\ 
\hline
\textbf{Intrusion}:\par{}\textit{actions that disturb one's solitude in physical space } & AI \textbf{\textit{exacerbates}} the scale and ubiquity of intrusion risks via enabling centralized and/or ubiquitous surveillance infrastructures. & Ring, a smart doorbell that enables homeowners to monitor activities and conversations near where the doorbell is installed has raised concern due to \textit{``the devices' excessive ability''} to capture data of an individual's neighbors \cite{milmo_amazon_2021}. \\ 
\hline
\multicolumn{1}{>{\hspace{0pt}}m{0.205\linewidth}}{} & \multicolumn{1}{>{\hspace{0pt}}m{0.285\linewidth}}{} & \multicolumn{1}{>{\hspace{0pt}}m{0.43\linewidth}}{} \\
\multicolumn{1}{>{\hspace{0pt}}m{0.205\linewidth}}{} & \multicolumn{1}{>{\hspace{0pt}}m{0.285\linewidth}}{} & \multicolumn{1}{>{\hspace{0pt}}m{0.43\linewidth}}{}
\end{longtable}
\vspace{-10mm}

\begin{figure*}[]
\centering
\includegraphics[width=0.75\columnwidth]{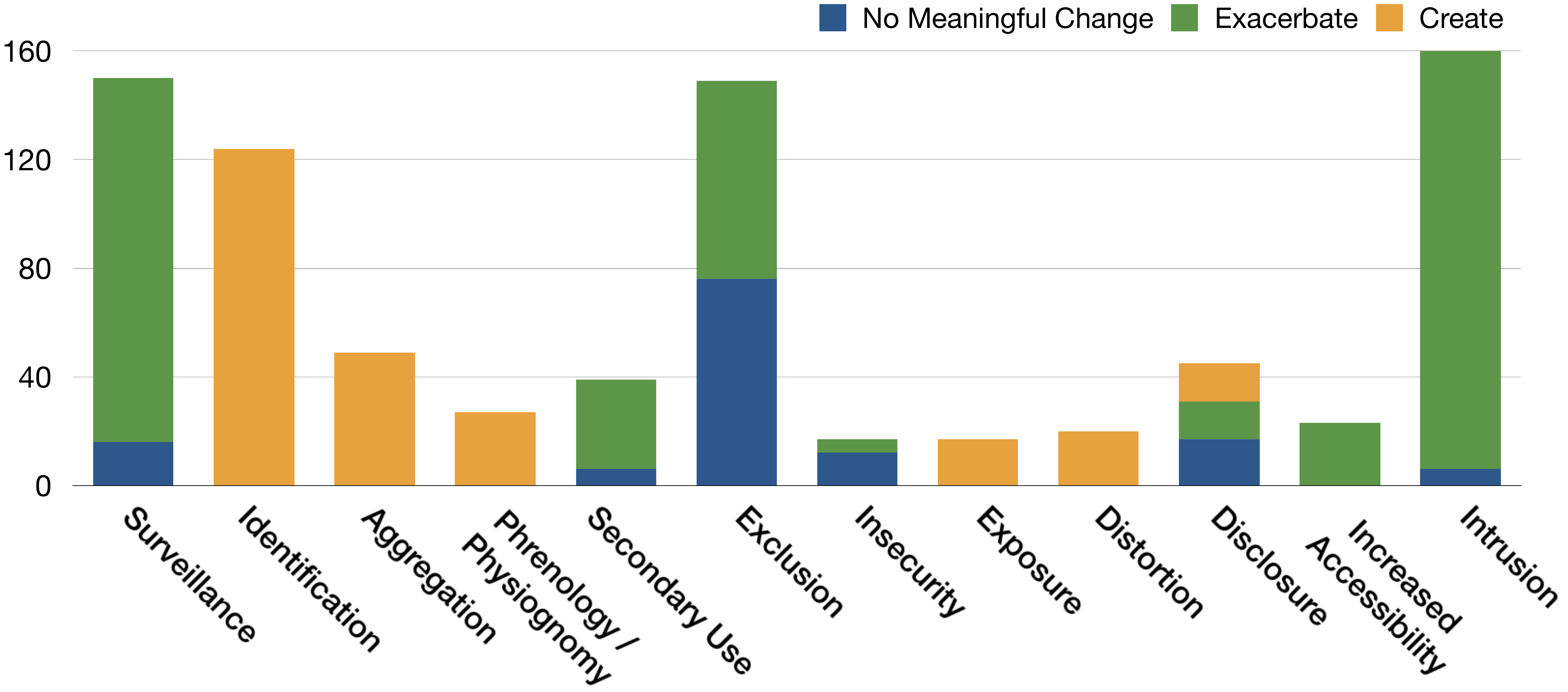}
\vspace{-2mm}
\caption{The distribution of each privacy risk we identified as not meaningfully changed, exacerbated, or created by AI. Note that one AI incident can involve multiple types of privacy risks.}
\label{fig:distribution}
\vspace{-3mm}
\end{figure*}

We introduce a taxonomy of AI privacy risks: i.e., privacy risks that are created and/or exacerbated by the incorporation of AI technologies into products and services. In short, we found that AI technologies create new instantiations of the privacy risks in Solove's taxonomy \cite{solove_taxonomy_2006} (e.g., generative AI can create new types of distortion intrusions), create a new category of risk not captured by Solove's taxonomy (e.g., resurging phrenology/physiognomy), and exacerbate many of the risks highlighted by Solove's taxonomy (e.g., AI technologies can more robustly identify individuals from low fidelity data sources) (see Figure~\ref{fig:distribution}). 

We discuss these AI-created and exacerbated risks below as they relate to data collection, processing, dissemination, and invasion (see Figure \ref{fig:model}).
Overall, we found that of the \totalcase{} incidents from the AIAAIC database that involve privacy risks, the AI technology implicated in the incident either created or exacerbated the described privacy risks 298 times (92.8\%), suggesting that the unique capabilities and/or requirements of AI do appear to meaningfully change privacy risks and that AI-specific privacy guidance may be necessary for practitioners.

\subsection{Data collection risks}
Data collection risks ``create disruption based on the process of data gathering'' \cite{solove_taxonomy_2006}.
Recent advances in AI/ML have been fueled by the collection of vast amounts of personal data. Solove further identifies surveillance as a risk that pertains to AI technology. 
AI technologies might \textit{create} data collection risks if the AI technology provides functionality that enables the collection of previously inaccessible data; they \textit{exacerbate} data collection risks when data is collected specifically for the development of an AI/ML system, or if AI technologies facilitate the data collection process in a manner that increases the scope of the risk.
In our analysis, we found incidents of AI exacerbating \textbf{surveillance} risks, but not of creating new such risks. 

\subsubsection{Surveillance (150/\totalcase{})}
Surveillance refers to watching, listening to, or recording an individual’s activities \cite{solove_taxonomy_2006}.
Surveillance risks long pre-date modern advances in AI.
AI technologies do not always meaningfully change surveillance (16/150), i.e., when end-users feed their own personal data to access the utility offered by AI, such as by uploading videos to capture body movement or estimate car speed.
Nevertheless, owing to the never-ending need for personal data to train and deploy effective machine learning models, we identified two ways AI technologies can exacerbate surveillance risks: i.e., by increasing the scale and ubiquity of personal data collected.

\textit{AI enhances the scale of surveillance (32/150)} by enabling linking across a diversity of sources, and increasing the quantity of collected personal data.

Where applicable, real-world models collect data from different sources to enrich datasets. 
We found that multi-faceted, high-fidelity data can exacerbate risks involving surveillance in the physical world.
One example comes from a predictive policing platform deployed in Xinjiang, China. The system \textit{``collects [individual's] information from a variety of sources including CCTV cameras and Wi-Fi sniffers, as well as existing databases of health information, banking records, and family planning history''}  \cite{rajagopalan_china_2018}. This information was then used to identify persons and assess their activities in the real world. 
We also found incidents describing AI systems that collected an array of end-user behavioral data in the cyber world. 
For example, Gaggle, a student safety management tool, monitors students' digital footprints such as email accounts, online documents, internet usage, and social media accounts to assess and prevent violence and suicides \cite{beckett_under_2019}.

Additionally, as the amount of training data often has a direct impact on model performance, AI technologies can exacerbate surveillance risks by increasing the need for collecting large-scale personal data to train effective models. 
For example, the South Korean Ministry of Justice attempted to build a government system for screening and identifying travelers based on photos of over 100 million foreign nationals who entered the country through its airports \cite{fassler_south_2021}.
Without the promise of AI technologies to automatically sift through and make sense of these data, there would be little incentive to collect data of this scale.

\textit{AI technologies exacerbate the ubiquity of surveillance risks (102/150)} by using physical sensors and devices to collect information from environments.
For example, geolocation data from mobile devices were used to assess employee performance, raising concerns about employee tracking outside of work \cite{velazco_amazons_2021}. CCTV cameras have been used in applications to detect and prevent suicide attempts \cite{park_seoul_2021} or to detect security anomalies in physical spaces \cite{wakefield_neighbour_2021}, while also introducing bystander privacy risks and concerns \cite{marky_i_2020}. Microphones enable a responsive audio interface for virtual assistants, along with concerns of extensive audio data collection and eavesdropping by the service provider \cite{paul_google_2019}.

\subsection{Data processing risks}
Data processing risks result from the use, storage, and manipulation of personal data \cite{solove_taxonomy_2006}.
Solove identified five types of data processing risks: identification, aggregation, secondary use, exclusion, and insecurity. In our analysis, we found incidents pertaining to each of these risks, as well as an entirely new category of data processing risk: \textbf{phrenology/physiognomy} risk, which is created by AI technologies by correlating arbitrary inputs and outputs.
We also found that AI technologies create new types of \textbf{identification} and \textbf{aggregation} risks (e.g., by operating on low-quality data;
and by forecasting future events), and exacerbate \textbf{secondary use}, \textbf{exclusion}, and \textbf{insecurity} risks (e.g., by re-purposing foundation models; by training models on datasets containing content obtained without consent; and by introducing new security vulnerabilities due to the use of AI).

\subsubsection{Identification (124/\totalcase{})}
\label{sec:identification}
Identification refers to linking specific data points to an individual’s identity \cite{solove_taxonomy_2006}.
These risks are commonplace even without AI; for example, users may be manually tagged in photos, or manually identified in CCTV video feeds.
AI technologies, however, allow for automated identity linking across a variety of data sources, including images, audio, and biometrics.
We found that AI technologies entail new types of identification risks with respect to scale, latency, robustness, and ubiquity.

\textit{AI technologies enabled automated identification at scale (20/124).} One example is Facebook's now-disabled Tag Suggestions product, through which Facebook demonstrated its ability to automatically identify individuals from uploaded photos. When this feature was in use, Facebook had 1.4 billion daily active users\footnote{\url{https://investor.fb.com/home/default.aspx}}; still, \textit{``any time someone uploads a photo that includes what Facebook thinks is your face, you'll be notified even if you weren't tagged''} \cite{simonite_facebook_2017}. 

\textit{AI technologies allow identification risks to occur more quickly, in nigh real-time (24/124)}, once the models are trained.
For example, in 2019, the Italian government was on the verge of implementing a real-time facial recognition system across football stadiums that ``prevent individuals who are banned from sports competitions from entering stadiums.'' It also picked up audiences' ``racist conversations'' to alert law enforcement authorities to the presence of racist fans \cite{spielkamp_automating_2019}.

In addition, \textit{AI technologies allow for robust identification even with low-quality data (7/124).} Clearview AI, a facial recognition application that aids U.S. law enforcement in identifying wanted individuals, claims to be able to identify people under a range of obfuscation conditions: \textit{``[a] person can be wearing a hat or glasses, or it can be a profile shot or partial view of their face''} \cite{hill_secretive_2020}. Similarly, models trained on Simulated Masked Face Recognition Dataset (SMFRD)\footnote{\url{https://github.com/X-zhangyang/Real-World-Masked-Face-Dataset\#download-datasets}} are capable of identifying persons with a mask on, \textit{``violating the privacy of those who wish to conceal their face''} \cite{wiggers_ai_2021}.

Finally, \textit{AI technologies enable ubiquity identification risks in situated physical environments (73/124)} like public places (e.g., \cite{mozur_inside_2018}), stores (e.g., \cite{hodgson_fast-food_2019}), and classrooms (e.g.,\cite{reuters_china_2018}). For example, XPeng Motors, a Chinese electric vehicle firm, was reported for using facial recognition-embedded cameras in their stores to collect biometric data of customers \cite{global_times_xpeng_2021}. 

\subsubsection{Aggregation (49/321)}
\label{result:aggregation}
Aggregation risks refer to combining various pieces of data about a person to make inferences beyond what is explicitly captured in those data \cite{solove_taxonomy_2006}.
These risks can occur without AI through manual analysis, but AI technologies greatly facilitate these inferences at scale, identified as a future trend by Solove: \textit{``the data gathered about people is significantly more extensive, the process of combining it is much easier, and the computer technologies to analyze it are more sophisticated and powerful''} \cite{solove_taxonomy_2006}.
Similar to identification risks, we found that AI technologies create new types of aggregation risks owing to their scale, latency, ubiquity, and their ability to forecast end-user behavior and infer end-user attributes.

One of the unique strengths of AI systems is that they automate complex processes into simple programs that overcome human limitations. While controversial, many public sectors still utilize algorithmic tools in high-stake contexts such as social work \cite{kawakami_why_2022} and services for the unhoused \cite{kuo_understanding_2023} to prioritize limited resources.
To that end, \textit{AI technologies create aggregation at scale (23/49)} by processing vast amounts of personal data to infer invasive things about individuals not explicit in those data. For example, an AI start-up created a service that assesses a prospective babysitter's likelihood to engage in risky behaviors such as drug abuse and bullying by \textit{``scan[ning] ... thousands of Facebook, Twitter and Instagram posts''} \cite{harwell_wanted_2018}.

\textit{AI technologies perform complicated inferencing tasks nigh instantly (11/49).} Technologies have been developed to estimate employee performance in-the-moment \cite{wakefield_amazon_2021},
and to forecast what one might write in emails \cite{tsukayama_gmails_2021}. AI technologies have also been developed to predict when end-users might be ovulating \cite{hunter_roe_2022}, and their moment-to-moment risk of committing suicide \cite{beckett_under_2019}. 

AI technologies can make physical objects and environments smarter and more responsive, \textit{enabling ubiquitous aggregation risks (5/49).} Smart home devices, for example, allow for automated control of home appliances, dynamic temperature control to strike an optimal balance between energy consumption and comfort, and voice user interfaces \cite{alam_review_2012}. 
These features require AI technologies to continuously monitor data streamed from physical sensors, creating new aggregation risks in situated environments. For example, smart speaker microphone feeds have been used to infer who is present in a room, who is speaking, and other information that can be algorithmically inferred from voice data \cite{noauthor_amazon_2018}. 

Finally, AI technologies enable forecasting future behaviors and states based on historical data. This forecasting can be used, for example, to help proactively identify health risks, plan optimal routes to avoid predictable traffic, and estimate retirement savings. These capabilities of AI, however, also \textit{create a new type of predictive aggregation risk (10/49).} 
For example, in 2018, Argentina's government deployed an AI model that predicted teen pregnancy in low-income areas from their first name, last name, and address \cite{jemio_case_2022}.
AI has also been used for crime prediction.
For example, in 2018, law enforcement in the United Kingdom aimed to predict serious violent crime using AI based on \textit{``records of people being stopped and searched and logs of crimes committed''} \cite{baraniuk_exclusive_2018}.

\subsubsection{Phrenology / Physiognomy (27/321)}
\label{result:p/p}
Phrenology and Physiognomy are debunked pseudosciences that postulate that it is possible to make reliable inferences about a person's personality, character, or predispositions from an analysis of their \textit{outer appearance} and/or \textit{physical characteristics} \cite{noauthor_face_2012}. 
Beyond the baseless prediction made from historical data streams discussed in Aggregation risks (Section \ref{result:aggregation}), phrenology/physiognomy risks pose unique downstream privacy harms distinct from aggregation risks: whereas aggregation risks primarily arise from the collection and combination of disparate pieces of information to make deductive inferences about individuals, phrenology/physiognomy risks introduce new and unfounded inferences about an individual's internal characteristics (e.g., their preferences and proclivities).
Moreover, while aggregation risks generally come from the combination of factual and observable data streams over which users can have some awareness and control (e.g., purchasing habits), phrenology/physiognomy risks arise from making inferences over physical characteristics over which users have no control.
Moreover, beyond the harm to the individual, there is also a broader societal harm: prior work has warned that irresponsible use of AI classification models could usher in a revival of these pseudosciences \cite{arcas_physiognomys_2017,stark2021physiognomic} by, e.g., motivating surveillance institutions to train AI models to make spurious inferences about a person's preferences, personality, and character from inputs that capture their outer appearance.
Our analysis reveals that AI technologies are indeed being used in this way, resulting in a new category of privacy risk not captured by Solove's initial taxonomy.
We define phrenology/physiognomy risks as the use of AI to infer personality, social, and emotional attributes about an individual from their physical attributes.
This risk stems from AI's ability to learn correlations between arbitrary inputs (e.g., images, voices) and outputs (e.g., one's demographic information).



Some models aim to infer preferences, like sexual orientation. For example, `Gaydar' is an AI sexual orientation prediction model that ``distinguishes between gay or straight people'' based on their photos \cite{levin_new_2017}. 
Researchers have also used AI to predict ``criminality'' --- i.e., whether someone is a criminal --- from facial images \cite{wu2016automated}.
Outside of the problematic assumptions of these models (i.e., that sexual orientation and criminality can be inferred from photos), this research raises concerns about the potential for harm and misuse of AI models to infer and disseminate information about individuals without consent \cite{levin_new_2017}.
AI technologies have also been used to predict other personal information such as one's name \cite{brewster_2_2021}, age \cite{press__checked_2015}, and ethnicity \cite{rollet_prc_2021} based on facial characteristics.

Other models aim to predict a person's mental and emotional state based on their images. For example, teaching tools devised by Class Technologies estimate students' engagement from their facial expressions without students' consent \cite{kaye_class_2022}. Still other models scrutinize vocal attributes to predict an individual's trustworthiness. For instance, the AI system DeepScore captures and assesses voice data to predict deceptiveness, and
has been utilized by health insurance and money lending platforms to select low-risk clients \cite{feathers_this_2021}.




\subsubsection{Secondary Use (39/\totalcase{})}
Secondary use encompasses the use of personal data collected for one purpose for a different purpose without end-user consent \cite{solove_taxonomy_2006}. In AI technologies, this risk is mostly associated with data practices for training data. 
AI does not always change secondary use risks (6/39). For example, Luca, an app that was used for contact tracing during the COVID-19 pandemic in Germany, was found to re-purpose personal data, such as location data, to support law enforcement by \textit{``tracking down witnesses to a potential crime''} \cite{pannett_german_2022}: but the risk described here would have been just as salient even without the use of AI.
Nevertheless, many common practices to train AI/ML models more effectively can exacerbate secondary use. 
In our dataset, we identified two AI-exacerbated secondary use risks: creating new AI capabilities with collected personal data, and (re)creating models from a public dataset.


When data collectors have already built models using personal data, they may be tempted to expand the models by creating additional features and capabilities, which can be unanticipated for end-users (22/39). 
For example, OkCupid, a dating site that matches users using an \textit{``one-of-a-kind algorithm''}\footnote{\url{https://www.okcupid.com/about}}, was found to contact an AI startup, Clarifai, \textit{``about collaborating to determine if they could build unbiased A.I. and facial recognition technology,''} and that \textit{``Clarifai used the images from OkCupid to build a service that could identify the age, sex and race of detected faces''} \cite{metz_facial_2019}.

Secondary use risks can also be exacerbated when AI practitioners try to reuse pubic datasets to train models for purposes other than the original purpose for which those data were collected (11/39). 
For example, People in Photo Albums (PIPA)
is a facial photograph dataset created to \textit{``recogniz[e] peoples' identities in photo albums in an unconstrained setting''} \cite{Zhang2015BeyondFF}. Yet, the PIPA dataset has been used in research affiliated with military applications and companies like Facebook \cite{Exposing.ai,harvey_exposingai_2021}.
Similarly, the Diversity in Faces (DiF) dataset is a collection of annotations of one million facial images that was released by IBM in 2019 \cite{smith_ibm_2019}. The dataset was created to improve the research on fairness and accuracy of artificial intelligence face recognition systems across genders and skin colors. While it was not to be used for commercial purposes, Amazon and Microsoft were accused of using the dataset to \textit{``improve the accuracy of their facial recognition software''} \cite{anne_long_amazon_2021}.

\subsubsection{Exclusion (149/\totalcase{})}
Exclusion refers to the failure to provide end-users with notice and control over how their data is being used \cite{solove_taxonomy_2006}.
Even without AI, computing products can covertly process data without informing users. Thus, AI technologies do not meaningfully change exclusion risks when the risk is isolated to just the covert processing of personal data (76/149).
For example, a ``trustworthiness'' algorithm developed by a short-term homestay company covertly used publicly accessible social media posts to ascertain if a potential customer was trustworthy \cite{johnston_banned_2022}, but the use of AI in this case did not fundamentally change the privacy risk. 
We nevertheless found in our incident database that the requirements of AI technology \textit{can} exacerbate exclusion risks by incentivizing the collection of large, rich datasets of personal data without securing consent (73/149). 

For example, the Large-scale Artificial Intelligence Open Network (LAION) is a German non-profit organization that aims ``to make large-scale machine learning models, datasets and related code available to the general public.'' In 2022, they released a large-scale dataset LAION-5B \cite{schuhmann_laion-5b_2022}, the biggest openly accessible image-text dataset at the time\footnote{https://laion.ai/blog/laion-5b/}. These data have been used to train many other high-profile text-to-image models such as Stable Diffusion\footnote{https://stablediffusionweb.com/} and Google Imagen\footnote{https://imagen.research.google/}\cite{edwards_artist_2022}. 
However, a person found that her private medical photographs were referenced in the public image-text dataset. She suspected that \textit{``someone stole the image from my deceased doctor's files and it ended up somewhere online, and then it was scraped into this dataset''} \cite{edwards_artist_2022}. 
Other models were found to be trained on ``semi-public'' personal data that were scraped from places like online forums, dating sites, and social media without users' awareness and consent (e.g., \cite{hill_secretive_2020,zimmer_okcupid_2016,noauthor_englands_2019}).
For example, Clearview AI built a private face recognition model trained on three billion photos that were \textit{``scraped from Facebook, YouTube, Venmo and millions of other websites''} \cite{mcdonald_clearviews_2020}. 

Prior work has shown that it can be challenging to ensure agency to any individual over their data regarding how data they have shared online can and cannot be used by such models \cite{obar2020sunlight}, and that it can be deliberately made complex for individuals to remove their data from the dataset \cite{birhane_multimodal_2021}. Additionally, when commercial AI models are ``black boxes,'' the general public has no means to audit how personal data is used by AI (e.g., Clearview AI). Finally, ``algorithmic inclusion'' --- i.e., ensuring that everyone is included in a system --- is often seen as a more desirable way to build AI systems in the context of AI ethics. 
These ``inclusive AI'' approaches, however, need to be balanced against exclusion-based privacy risks \cite{albert_algorithmic_2022,Andrus2022DemographicReliantAF}: when more people's data are captured to build inclusive systems, those people may be subject to increased exclusion risk if their data is collected without adequate consent and control. 

\subsubsection{Insecurity (17/\totalcase{})}
Insecurity refers to carelessness in protecting collected personal data from leaks and improper access due to faulty data storage and data practices \cite{solove_taxonomy_2006}. Products and services that include AI are subject to many of the same insecurity risks that result from poor operational security, unrelated to the capabilities and data requirements of AI (12/17). For example, our dataset includes a data breach where attackers hacked into Verkada, a security startup that provides cloud-based security cameras with face recognition. This gave the attackers access to cameras that \textit{``are capable of identifying particular people across time by detecting their faces, and are also capable of filtering individuals by their gender, the color of their clothes, and other attributes''} \cite{cox_hacked_2021,turton_hackers_2021}.
These operational security mistakes are not unique to or exacerbated by AI technologies, even though the AI-enabled products and services that are hacked afford attackers access to compromised data that would otherwise not be accessible. We did, however, find instances in which the capabilities and/or data requirements of AI technologies directly exacerbated insecurity risks (5/17).

Sometimes AI technology can compromise end-user privacy in order to enable AI utility. For example, Allo, a messaging app that Google first launched in 2017, included an AI virtual assistant and automatic replies.
The messenger was not end-to-end encrypted, allowing for AI models developed by Google to ``read'' users' chat content and personalize services for them \cite{gebhart_googles_2016}. 

We also found cases where AI technologies unexpectedly reveal the personal data on which they were trained. For example, Lee Luda, a chatbot trained on real-world text conversations, was found to expose the names, nicknames, and home addresses of the users whose data on which it was trained \cite{jang_south_2021}. Similarly, services that use generative AI models to create realistic but fake human faces, have been shown to be able to reconstruct the raw personal data on which the models were trained \cite{webster_this_2021}.

Additional vulnerabilities can be introduced through the infrastructural data requirements entailed by AI technologies. For example, converting raw data into training-ready labeled data can require the exposure of raw personal data to human annotators. For example, iRobot hired gig workers to annotate audio, photo, and video data captured by their household robots to train AI models. However, some of these raw and sensitive photos were leaked online by the gig workers \cite{guo_roomba_2022}. Cases like this illustrate how AI can blur the boundary between data \textit{processing} risks and data \textit{dissemination} risks --- sometimes, the act of processing data through AI requires dissemination.

\subsection{Data dissemination risks}
Data dissemination threats result when personal information is revealed or shared by data collectors to third-parties \cite{solove_taxonomy_2006}.
AI technologies \textit{create} new data dissemination risks by enabling new ways of revealing and spreading personal data; they also \textit{exacerbate} data dissemination risks by increasing the scale and the frequency of the dissemination.

In our analysis, we found that AI technologies create new types of \textbf{exposure}, \textbf{distortion}, and \textbf{disclosure} risks (e.g., by reconstructing redacted content; by generating a realistic fake video of an individual; and by sharing AI-derived sensitive information about individuals with third-parties).
We also found cases in which AI technologies exacerbated known \textbf{disclosure} risk (e.g., by sharing large-scale user data to third-parties to train models), and \textbf{increased accessibility} risk (e.g., by open-sourcing large-scale benchmark datasets containing user data).

\subsubsection{Exposure (17/\totalcase{})}
Exposure risks encompass revealing sensitive private information that people view as deeply primordial that we have been socialized into concealing \cite{solove_taxonomy_2006}.
Traditionally, these risks arise when an individual's private activities are recorded and disseminated to others without consent.
AI technologies can create new types of exposure risks via generative techniques that can create, reconstruct, manipulate content (i.e., deepfake techniques)
(10/17) and expose inferred sensitive end-user attributes predicted by AI/ML
(e.g., one's interests \cite{levin_new_2017}) (7/17). 

Specifically, we found that AI can create new types of exposure risks by reconstructing censored or redacted content. For example, generative adversarial networks (e.g., TecoGAN \cite{chu_learning_2020}) have been used to clarify images of censored genitalia \cite{montgomery_man_2021}, and to ``undress'' people to create pornographic images without consent \cite{burgess_biggest_2021}. 
Deepfake applications such as DeepFaceLive\footnote{\url{https://github.com/iperov/DeepFaceLive}} or DeepFaceLab\footnote{\url{https://github.com/iperov/DeepFaceLab}} can be made to morph a non-consenting subject's face into pornographic videos.
These deepfake technologies have been used to facilitate mass dog-piling and online harassment \cite{ayyub_india_2018} and to create illegal online pornography businesses \cite{noauthor_deepfake_2021}.

In our analysis, we also found that AI technologies create new risks that expose sensitive data, preferences, and intentions inferred by AI/ML. 
For instance, Flo,
an app that tracks menstruation and ovulation, forecasts its users' menstrual cycle and ovulation.
Despite promising to maintain the privacy of personal data, Flo allegedly shared customers' menstrual timing and intention to get pregnant with third-parties like Facebook \cite{schechner_you_2019}.
AI can also be built to proactively disseminate incriminating information about individuals to the public.
In Shenzhen, China, a system was implemented to detect jaywalking and other offenses captured by cameras.
The system identifies offenders and displays their photographs, names, and social identification numbers on LED screens placed at road junctions \cite{xu_chinese_2018}.

\subsubsection{Distortion (20/\totalcase{})}
Distortion refers to disseminating false or misleading information about people 
\cite{solove_taxonomy_2006}.
Distortion risks are analogous to slander or libel, and have existed well before modern advances in AI.
However, we found that AI technologies can create new types of distortion risks by exploiting others' identities to generate realistic fake images and audio that humans have difficulty discerning as fake \cite{vincent_binance_2022,nightingale_ai-synthesized_2022}. 

Some models can generate realistic audio of individuals. For example, Prime Voice AI,
a text-to-voice generator, was misused to create the voices of celebrities to \textit{``make racist remarks about Alexandria Ocasio-Cortez (the US House representative)''}, and that the AI-generated clips \textit{``run the gamut from harmless, to violent, to transphobic, to homophobic, to racist.''} \cite{cox_ai-generated_2023, harrison_startup_2023}.
Other AI-created distortion risks are less egregious, but raise important questions about expectations around privacy in light of how generative AI can be used to simulate the likeness of those who have passed.
For example, the filmmaker of a documentary was revealed to be using deepfake technology to create scenes, with the likeness of an actor who had passed away, for lines \textit{``he wanted [Anthony] Bourdain's (the main character of the documentary) voice for but had no recordings of''} \cite{leon_roadrunner_2021}. 


\subsubsection{Disclosure (45/\totalcase{})}
\label{sec:disclosure}
Whereas distortion is the dissemination of false or misleading information, disclosure risks encompass the act of revealing and improperly sharing people's personal data
 \cite{solove_taxonomy_2006}.
Indeed, any computing product that collects and stores personal data can introduce disclosure risks. Our dataset includes cases where AI does not meaningfully change disclosure risks (17/45), such as sharing personal data with law enforcement or third-parties.
Nevertheless, AI technologies create new types of disclosure risks by being able to derive or infer additional information beyond what is explicitly captured in the raw data.
We also found AI technologies can exacerbate disclosure risks because the personal data used to train ML models are often shared with specific individuals or organizations.

Many of the disclosure risks we identified involved the creation of machine learning models that automatically infer undisclosed personal information about individuals (14/45). 
For example, the ``Safe City'' initiative in Myanmar used AI-infused cameras to identify faces and vehicle license plates in public places and alert authorities to individuals with criminal histories \cite{noauthor_myanmar_2021}.



AI technologies can also exacerbate disclosure risks when personal data is shared by organizations to train machine learning models (14/45). For example, 
the UK's National Health Service partnered with Google to share mental health records and HIV diagnoses of 1.6 million patients to develop a model for detecting acute kidney injury \cite{hodson_revealed_2016}.

\subsubsection{Increased Accessibility (23/\totalcase{})}
Increased accessibility refers to making it easier for a wider audience of people to access potentially sensitive information.
We found incidents in which AI technologies exacerbated the scale of this risk via the public sharing of large-scale datasets, containing personal information, for the use of building and improving AI/ML models. In the AI/ML community, it is common practice to leverage open-source benchmark datasets to train AI/ML models. This open-source data sharing enables transparency and public audits of AI research and development. However, publicizing datasets also enables anyone to collect large amounts of personal data that may have otherwise been private, access-controlled, or difficult to find.
For example, the ``OkCupid dataset'' contained data of almost seventy thousand users from the dating site OkCupid. The dataset contained personal information such as users' location, demographics, sexual preferences, and drug use. It was uploaded to Open Science Framework, a website that helps researchers to open source datasets and research software, to facilitate research on modeling dating behaviors \cite{woollacott_70000_2016}.



\subsection{Invasion risks}
The final top-level category of privacy risk Solove outlined, Invasion, can be understood as the unwanted encroachment into an individual's personal space, choices, or activities \cite{solove_taxonomy_2006}. 
Solove placed two sub-categories under invasion: intrusion and decisional interference. We found incidents where AI technologies exacerbated intrusion risks, in particular.


\subsubsection{Intrusion (160/\totalcase{})}
Intrusion risks encompass actions that disturb one's solitude in physical space \cite{solove_taxonomy_2006}.
For six of the 160 intrusion incidents we identified, we noted that the AI technologies described in the incident did not fundamentally change the risk described in the incident: the intrusion would have remained as described even without the capabilities and/or requirements of AI. One example is the use of digital screens in stores to show customers personalized ads \cite{meyersohn_walgreens_2022}: the intrusion would remain, even if the system did not use AI.
However, we identified two ways AI can exacerbate intrusion risks that increase their scale and ubiquity.

The capabilities of AI technologies (e.g., to identify a person and detect behaviors) \textit{enable a centralized surveillance infrastructure that creates large-scale intrusion risks (113/160)}; the requirements of AI (e.g., access to vast troves of data and GPU servers) necessitate this infrastructure. 
For example, Pharmaceutical University in Nanjing, China, implemented a recognition system at various locations on campus to closely monitor students' attendance and learning behaviors \cite{tone_camera_2019, zhang_chinese_2019}.
Similarly, employers are increasingly incorporating AI-infused workplace monitoring technologies that collect data from employees' smartwatches \cite{the_local_spains_2020} and computer webcams \cite{walker_call_2021} to track their performance, absence, and time-on-task.

The capabilities of AI can also \textit{turn everyday products (e.g., doorbells, wristbands) into powerful nodes in a ubiquitous surveillance infrastructure (41/160).}
For example, Ring, a smart doorbell that enables homeowners to monitor activities and conversations near where the doorbell is installed, has raised concern due to ``the device's excessive ability'' to capture data of an individual's neighbors \cite{milmo_amazon_2021}. 
Similarly, Amazon's Halo fitness tracker uses AI to analyze a user's conversations to highlight when and how often that user spoke in a manner that was indicative of their being ``happy, discouraged, or skeptical'' \cite{ohara_amazon_2021}.

\section{Discussion}
Our findings demonstrate the many ways modern advances in AI meaningfully change privacy risks relative to how we conceived of privacy risks prior to these advances, as captured by Solove's widely cited taxonomy of privacy \cite{solove_taxonomy_2006}. Across the \totalcase{} AI privacy incidents we analyzed, roughly 7\% of the cases did not involve privacy risks that were created or exacerbated by AI. For example, we encountered instances where a product that happened to include AI was subject to a data breach in which users' personal data was compromised \cite{abrams_proctoru_2020}. Nevertheless, in approximately 93\% of the cases we analyzed, the unique capabilities and data requirements of the AI technologies involved in the incident either created a new type of privacy risk, or exacerbated a known risk.

We found that the unique capabilities of AI create new types of privacy risks. For example, AI creates new data processing risks in its ability to identify the activity of individuals even with low-quality data, and in its ability to forecast future outcomes. AI creates a new category of phrenology/physiognomy risks by enabling the creation of spurious classifiers correlating physical attributes with social, emotional, and personality traits. AI creates new types of data dissemination risks in its ability to generate human-like media, e.g., by generating a realistic fake video of an individual. We also found that the data requirements of AI exacerbate privacy risks we have grappled with for decades. For example, AI technologies can lead to more pervasive, larger scale surveillance than before; exacerbate secondary use, exclusion, insecurity, disclosure, and increased accessibility risks in the processing and disseminating of personal data; and, increase the ways in which computing can intrude upon people's personal space.

Equipped with the knowledge of how AI \textit{has} changed privacy risks, we first discuss how the current AI/ML methods fall short and only address a subset of the AI privacy risks identified in our taxonomy (Section~\ref{sec:privacy-preserving AI}). Then, we present our taxonomy as a living structure that can be expanded with risks documented by Solove's original taxonomy \cite{solove_taxonomy_2006} in cases where we did not find matching incidents in our incident database (Section~\ref{sec:theoretical ai privacy harms}). In theory, future advances in and/or the use of AI may entail risks in these categories, so it is worth discussing them as privacy risks that AI may change in the future.
Moreover, we discuss a number of ways we expect this taxonomy might be useful for both future research and practice (Sections~\ref{sec:future work ai privacy guidance} and \ref{sec:future work living taxonomy}).

\subsection{Charting the design space for privacy-preserving AI/ML work}
\label{sec:privacy-preserving AI}
Our findings broaden the design space for privacy-preserving AI and ML. For example, a recent meta-review of HAI principles and guidelines argues that privacy in ML-driven systems centers around the protection, control, and agency over personal data \cite{yurrita_towards_2022}. Based on our findings, these criteria only consider a small subset of the AI privacy risks we identified: they consider some --- but not all --- of the data collection and processing risks exacerbated by AI, and do not at all consider the data processing and dissemination risks newly created by AI. In this section, we provide an overview of how the existing tools and approaches, that aim to help practitioners build privacy-preserving AI systems \cite{yurrita_towards_2022,meurisch_data_2022,wong_seeing_2023}, fall short of effectively identifying and addressing many AI privacy risks.

\paragraph{Differential Privacy and Federated Learning}
Differential Privacy (DP) \cite{nasr_machine_2018} and Federated Learning (FL) \cite{li_federated_2020} are commonly thought of as approaches to ``privacy-preserving'' machine learning where 1) the model output is insensitive to the presence or absence of data on an individual in a dataset, and 2) the model provider only learns and improves the model in an aggregated manner. Tools such as Diffprivlib\footnote{\url{https://github.com/IBM/differential-privacy-library}} \cite{diffprivlib} and IBM Federated Learning\footnote{\url{https://github.com/IBM/federated-learning-lib}} \cite{diffprivlib} have been used by practitioners to implement DP and FL into their ML products.
When training an ML model, however, these approaches only apply to some data processing risks --- e.g., so that the model can not be used to re-identify data of individuals from the model outputs --- and not the full range of risks we discuss in our taxonomy. Owing to these shortcomings, organizations that commonly advocate for end-user privacy rights, like the Electronic Frontier Foundation (EFF), have argued against the use of these approaches when they are used as stand-ins for stronger privacy protections (e.g., as in the case of Google's attempt to replace third-party browser cookies with ``Federated Learning of Cohorts'') \cite{cyphers_googles_2021}. For example, the ``criminality classifier'' that takes in photos of people's faces and claims to predict their likelihood to be a criminal \cite{wu2016automated} could be built with a federated learning architecture. Doing so would not address the physiognomy risk inherent to the idea itself, nor the exclusion and disclosure risks arising from how the data is collected and the inferences shared without consent.

\paragraph{Data Privacy Auditing}
Prior work has created data auditing tools, such as the Privacy Meter\footnote{\url{https://github.com/privacytrustlab/ml_privacy_meter}} \cite{murakonda2020ml}, to help practitioners conduct privacy impact assessments on ML models. Doing so allows practitioners to quantify some privacy risks (e.g., membership inference attacks). However, because the Privacy Meter must be applied \textit{after} the model is trained, it is inherently limited in its ability to mitigate against the risks that arise in the data collection and processing phases of work.
In addition, similar to DP and FL, this approach takes a limited view of privacy and only applies to specific data processing risks --- e.g., aggregation risks that arise from collective sensitive personal data in the training data.


\paragraph{Ethics Checklists and Toolkits}
Prior work in AI ethics has introduced many toolkits to support practitioners in ethical AI development \cite{wong_seeing_2023}, some of which also surface privacy risks. For example, Microsoft's Harms Modeling\footnote{\url{https://learn.microsoft.com/en-us/azure/architecture/guide/responsible-innovation/harms-modeling/}} is an activity that includes design exercises and worksheets that help \textit{``evaluate potential ways the use of a technology you are building could result in negative outcomes for people and society,''} including potential privacy risks. AI ethics checklists such as Deon\footnote{\url{https://github.com/drivendataorg/deon}} allow practitioners to \textit{``add an ethics checklist to [their] data science projects,''} which include questions that make practitioners reflect on the collection, storage, and analysis of data containing PII (personally identifiable information). These checklists and toolkits could help practitioners consider a broader range of privacy risks described in our taxonomy (e.g., data collection and dissemination risks). However, these tools approach privacy risks monolithically, and at a high-level (e.g., privacy loss, PII exposure); they provide little guidance to practitioners to consider privacy risks newly created and/or exacerbated by AI (e.g., physiognomy, distortion risks). In other words, the use of such tools relies on practitioners' individual awareness of AI privacy risks, which prior work has identified as a key barrier to AI privacy work \cite{lee2024}.

Note that \textit{all} of these approaches have value and we are not suggesting that they not be used. Rather, we caution against rhetoric that it is possible to create ``privacy-preserving'' AI/ML technologies using \textit{only} these approaches.

\subsubsection{Future Work: Creating AI-specific privacy guidance}
\label{sec:future work ai privacy guidance}
Given that our findings show that AI creates new types of privacy risks and exacerbates existing ones, and that current privacy-preserving AI/ML methods fall short of identifying and addressing many of these risks,
there is a need for future work to fill the gap of mitigating privacy risks created and exacerbated by AI.
Specifically, our taxonomy opens up a new design space for privacy-preserving AI/ML tools that aim to raise practitioners' awareness of utility-intrusiveness trade-offs of their AI product ideas (e.g., \cite{ernala_exploring_2021}). 
For example, prior work in other AI-adjacent fields, such as Robotics, has explored how to correlate desired robot function with a minimally-invasive set of sensors \cite{eick_enhancing_2020}.
In the broader context of implementing privacy and security in software products, prior work has found that practitioners still largely see privacy and security in products as an ``all or nothing'' notion such that privacy comes at the expense of other important objectives \cite{tahaei_privacy_2021,gutfleisch_how_2022}. 

Future work can explore incorporating our AI privacy taxonomy into harm-envisioning techniques, such as Consequence Scanning \cite{doteveryone_consequence_2020}, by providing AI privacy risk prompts to capture associated negative consequences holistically.
These techniques can help practitioners run lightweight privacy evaluations on AI product ideas, and help them balance the utility and intrusiveness of these products and services across design iterations.
With such a tool, we hypothesize that practitioners can better advocate and design for privacy in working contexts that may dissuade this work \cite{tahaei_privacy_2021,lee2024}.

Our taxonomy can also consolidate promising future research in foregrounding tensions across data pipelines, practices, and stakeholders (i.e., data subjects, data observers, data beneficiaries, and data victims).
By mirroring the first step in Rahwan's Society in the Loop framework \cite{rahwan_society---loop_2018}, AI practitioners can make concrete the envisioned value and the stakeholders of their proposed AI concepts. 
To assist in this process, future work can create artifacts that encourage practitioners to articulate the value proposition of their envisioned product.
Based on our taxonomy, then, it may be possible to mine our database for AI privacy incidents about products that are ``semantically'' similar based on an articulated value proposition.
By showing practitioners related AI privacy incidents, they might then be guided to reflect on the utility-intrusiveness trade-off of their envisioned AI product ideas: 
for whom that value is generated (i.e., data beneficiaries), whose data is processed to unlock that value (i.e., data subjects), who can be impacted by the data pipeline (i.e., data victims), and by which privacy risk (e.g., surveillance).


In practice, however, this type of early-stage discussion around AI utility and privacy risk can be challenging because: (i) practitioners do not necessarily understand the full potential and limitations of AI \cite{yildirim_how_2022}; (ii) privacy is often treated as compliance with general regulatory mandates rather than a product-specific design choice \cite{waldman_designing_2017}; and, (iii) practitioners do not have access to AI-specific tools that support their privacy work pertaining to the capabilities and requirements that AI brings to their products \cite{lee2024}. Accordingly, there is a need for a greater understanding of where such tools and artifacts might be effectively incorporated into practitioners' workflows.

\subsection{Theoretical extensions to the AI privacy risks taxonomy}
\label{sec:theoretical ai privacy harms}
We see our taxonomy as a living structure that helps scaffold the conversation about how advances in AI change privacy risks.
But just as the capabilities and requirements of AI may change with future advances, so too might AI privacy risks.
One way we might envision future AI privacy risks is by exploring the four subcategories of privacy risk in Solove's original taxonomy \cite{solove_taxonomy_2006} for which we did not find relevant incidents in our dataset: Interrogation, Blackmail, Breach of Confidentiality, and Decisional Interference.
In the future, we may observe incidents where advances in AI meaningfully change or exacerbate these risks as well.

\paragraph{Interrogation}
Interrogation risks encompass the covert collection of data while a subject is being actively questioned \cite{solove_taxonomy_2006}. 
For example, lie detector tests entail interrogation risks --- information beyond what an individual is saying is collected to assess the truthfulness of their words.
We can envision AI both creating and exacerbating interrogation risks.
Large Language Model-powered chatbots like ChatGPT, for example, could create new interrogation risks by imitating people and interacting with users in natural language, aiming to extract information from users.
AI-infused affective computing technologies could exacerbate interrogation risks (e.g., \cite{noriega_application_2020}): using these technologies, it may be possible to draw inferences about an individual's demeanor from verbal (e.g., language use, tone) and non-verbal (e.g., body language, eye movements) cues.

\paragraph{Blackmail}
Blackmail refers to coercing individuals by threatening to disclose private or sensitive information \cite{solove_taxonomy_2006}.
Generative AI technologies could create new instantiations of this risk by synthesizing fake but convincing content that may serve as evidence for blackmail. 
We already saw incidents where incriminating content was fabricated when describing the exposure and distortion risks in our taxonomy, but we did not see this fabricated content being used for blackmail in the incidents we analyzed.
Moreover, by automating the process of gathering and compromising information at scale, AI can also exacerbate blackmail risks.
As we have seen, ML algorithms can analyze vast datasets from social media, location services, and personal files to identify content that could be used as fodder for blackmail. 

\paragraph{Breach of Confidentiality}
Breach of Confidentiality refers to an interpersonal risk between two people where one party discloses something to the other in confidence, and the other party violates this confidence by sharing it with third-parties \cite{solove_taxonomy_2006}. 
AI technologies could exacerbate the scale of this risk by enabling conversational agents capable of gaining users' trust and guiding them to share sensitive information. 
For example, attackers can deploy such AI systems in high-stakes scenarios like healthcare and finance, and pose threats of breaching the confidentiality of the users by sharing the sensitive information they shared with the agent to third-parties.

\paragraph{Decisional Interference}
Decisional Interference concerns the unwanted influence over or constraint of an individual's choices or behavior by a third-party \cite{solove_taxonomy_2006}. 
Solove specifically focuses on the government as the relevant third-party, but private institutions and enterprises can also be culprits for this category of risk.
AI technologies can exacerbate decisional interference risks by enabling more personalized political propaganda (e.g., \cite{shapiro_measuring_2021}).
AI technologies might also exacerbate the scale of existing practices of online censorship toward political topics (e.g., \cite{bordelon_we_2023}).
Algorithms for personalized recommendation or persuasive technologies can also subtly guide user choices, sometimes in ways that align more with the goals of external entities (e.g., advertisers or political campaigns) than with the individual's own preferences or well-being. 

\subsubsection{Future Work: Creating a living taxonomy of AI privacy risks}
\label{sec:future work living taxonomy}
To our knowledge, our taxonomy is the first attempt to show how common AI requirements and capabilities map onto high-level privacy risks. As shown above, future AI privacy incidents can also expand the taxonomy.
In addition, future AI privacy incidents may create new categories of privacy risk that go beyond Solove's taxonomy (like the physiognomy risk we describe here). For example, many artists have been vocal about concerns about the theft of artistic style by generative AI \cite{knibbs_new_2023}. While these discussions currently center around notions of copyright and intellectual property, we can envision new types of privacy risk as well: e.g., artistic styles might contain personally identifiable or sensitive information.
We envision that our taxonomy can complement ongoing crowd-sourced efforts at curating and organizing AI incidents such as the AIAAIC \cite{pownall_ai_2023} and AIID\footnote{\url{https://incidentdatabase.ai/}} by providing a framework to formally synthesize and identify emerging privacy risks in AI incidents. 
With that in mind, the research team is building a website\footnote{The website will be available at \url{https://privacytaxonomy.ai/} and \url{https://aiprivacytaxonomy.com/}} to present our taxonomy of AI privacy risks, and is also planning to expand this website to collect and aggregate submissions of new incidents related to these risks.

To present the AI privacy taxonomy in forms useful to the HCI and AI communities, future work can take an iterative approach, grounded on practitioners' and academics' 
actual design and research needs, to model the translation function between AI technology ideas and potential risks to consider.
Indeed, envisioning with AI --- i.e., treating AI as a design material \cite{yang_investigating_2018,jansen_mix_2023,yildirim_how_2022,yildirim_creating_2023} --- is an open and active area of research.
Aligning with this line of research, future work can add to our taxonomy by systematizing AI capabilities and requirements, and the privacy risks they create and exacerbate, at a level of granularity that is useful for practitioners and researchers to ideate, communicate, and collaborate with product teams and stakeholders \cite{yildirim_how_2022,yildirim_creating_2023}.

\subsection{Limitations}
We consciously took an ``incident-based'' approach when constructing our taxonomy. There is a great deal of hype about what AI technologies can do, blurring the lines between speculation and reality \cite{kapoor_ai_2023}. The overabundance of speculative risks necessitated that we limit our consideration to those that journalists and the public-at-large have recognized as harmful as chronicled in the AIAAIC database.
With that in mind, our dataset should not be interpreted as inclusive and representative of every \textit{possible} privacy risk created or exacerbated by AI technologies: it is a repository of many privacy risks that have been realized in practice.

Our goal in creating this taxonomy was to codify AI privacy risks based on an accounting of documented, real-world risks.
To that end, AIAAIC is currently \textit{``the most comprehensive, detailed, and timely resource''}\footnote{\url{https://www.aiaaic.org/aiaaic-repository/about-the-aiaaic-repository}} that is openly accessible and has been used by the community as the source to synthesize the harms caused by AI functionality \cite{raji_fallacy_2022}. 
To mitigate the sampling bias introduced by our use of the AIAAIC, we tested the database's coverage by independently collecting a list of 15 AI privacy incidents from various sources, e.g., social media posts, literature. Of the 15 incidents we collected, 13 were also included in AIAAIC. For the two incidents that were not included, we found very similar incidents in the database --- i.e., similar privacy risks caused by the same technology (e.g., face recognition software) but of different products. As a comparison, we applied the same procedure to the AIID database and only found five incidents included.
Thus, we believe that AIAAIC currently provides a pool of AI privacy accidents comprehensive enough for our goal.

We acknowledge that there will be a growing number of AI incidents, and that there may be existing AI incidents that were not captured in our dataset. For example, prior work has surfaced how algorithmic recommender systems can amplify embarrassing exposures through online social networks \cite{choi_embarrassing_2015}. 
Nevertheless, our taxonomy provides a solid foundation for understanding how the capabilities and requirements of AI change privacy risks.
Since we ground our taxonomy on Solove's taxonomy of privacy, which has remained highly influential and largely appropriate for nearly two decades, we are confident that our updated taxonomy can be flexibly adapted to encompass new risks if and as they are realized beyond academic inquiry.

Finally, we acknowledge that ``privacy'' is a broad and context-dependent concept that is susceptible to biased interpretation based on the research team's background. We are an interdisciplinary research team with diverse expertise across HCI, AI, security and privacy, policy, and design. We mitigated the potential for bias by: (i) building our taxonomy on top of Solove's existing and widely accepted taxonomy; (ii) ensuring that multiple coders independently agreed on the risks entailed (or not) by a specific incident; and, (iii) dutifully analyzing \textit{all} incidents, in the AIAAIC database, that were independently characterized by people outside of our research as being privacy-pertinent.


\section{Conclusion}
In this paper, we conducted a systematic analysis of documented incidents of AI privacy risks to answer the question: How do modern advances in AI and ML change privacy risks?
Our taxonomy, constructed from a corpus of \totalcase{} documented AI privacy incidents, reveals that while the incorporation of AI technologies into products does not \textit{necessarily} change the privacy risks those products might entail, it often does. Our taxonomy reveals that AI can create new types of privacy risks when processing and disseminating end-user data. We showed, for example, that the unique capabilities of AI technologies (e.g., the ability to generate realistic but fake images) also create new types of privacy risks (e.g., exposure risks from deepfake pornography \cite{ayyub_india_2018}). 
The taxonomy also reveals that the data requirements of AI technologies can exacerbate known privacy risks.
For example, owing to the unique ability of AI to automatically identify individuals from low-fidelity images, governments are more motivated to capture facial images of all passengers that pass through major transportation hubs (e.g., \cite{fassler_south_2021}).
Our work suggests that AI-specific design guidance is needed for practitioners to negotiate the utility-intrusiveness trade-offs of AI-powered user experiences, and that many existing approaches to privacy-preserving machine learning (e.g., federated learning \cite{li_federated_2020}) address only a small subset of the unique privacy risks entailed by AI technologies.


\begin{acks}
This work was generously funded, in part, by NSF SaTC grants \#2316768 and \#2126066. We thank our anonymous reviewers for their helpful comments in revising this paper. We also thank the contributors and maintainers of the AIAAIC incident database.  
\end{acks}

\bibliographystyle{ACM-Reference-Format}
\bibliography{ptd,ptd_zotero,news}

\appendix
\section{Appendix}

\begin{table}[h]
\centering
\captionof{table}{Cohen's Kappa for each privacy risk.}
\label{tab:irr}
\resizebox{0.35\textwidth}{!}{
\begin{tabular}{@{}cc@{}}
    \toprule
    \textbf{Privacy Risk} & \textbf{Cohen's Kappa} \\ \midrule
    Surveillance               & 0.88           \\
    Identification             & 0.94           \\
    Secondary Use              & 0.84          \\
    Aggregation                & 0.87          \\
    Phrenology / Physiognomy    &   1 \\
    Exclusion                  & 0.90           \\
    Insecurity                 & 0.94           \\
    Exposure                   & 1              \\
    Disclosure                 & 1              \\
    Distortion                 & 1              \\
    Increased Accessibility   & 1               \\
    Intrusion                  & 0.94  \\
    \midrule
    Average                     & 0.94 \\
    
    \bottomrule
    \end{tabular}
    }
\end{table}

\end{document}